\documentclass[aps,prb,twocolumn,superscriptaddress]{revtex4-2}
\usepackage{graphicx}
\usepackage{dcolumn}
\usepackage{color}
\usepackage{bm}
\usepackage{amsmath}
\usepackage{mathtools}
\usepackage{float}

\newcommand{\cc}{\,cm$^{-3}$~}
\newcommand{\rxx}{$\rho_{xx}$}
\newcommand{\rxy}{$\rho_{xy}$}
\newcommand{\cxx}{$\sigma_{xx}$}
\newcommand{\cxy}{$\sigma_{xy}$}

\begin{document}

%\preprint{}

\title{Large linear magnetoresistance and evidence of degeneracy lifting of valence bands in rhombohedral phase of topological crystalline insulator SnTe}

\author{Sonali Baral}
\author{Mukesh Kumar Dasoundhi}
\author{Indu Rajput}
\author{Devendra Kumar}
\author{Archana Lakhani}
\email[Corresponding author: ]{archnalakhani@gmail.com}
\affiliation{UGC-DAE Consortium for Scientific Research, University Campus, Khandwa Road, Indore-452001, INDIA}

\date{\today}

\begin{abstract}
  We report a comprehensive magneto-transport study on single crystalline p-type topological crystalline insulator (TCI) SnTe, across the cubic-to-rhombohedral (\emph{R3m}) transition which occurs as a function of temperature. The electrical resistivity of a well-characterized SnTe crystal shows evidence for the cubic-to-rhombohedral structural transition at T$_s$ $\sim$~64\,K and a carrier density of $\sim$1.8$\times$10$^{20}$\cc at 77\,K. As a function of applied magnetic field perpendicular to the (100) plane, SnTe exhibits a large unsaturated linear magnetoresistance (LMR) reaching a value of 42\,\% at 5\,K and 8\,T. LMR is found to have a direct dependence on the mobility and a detailed analysis shows that it follows the classical Parish-Littlewood model of conductivity fluctuations arising from macroscopic inhomogeneity of tellurium interstitial atoms. We also observe SdH oscillations in the rhombohedral (\emph{R3m}) phase with a Berry phase of $\pi$ and significantly lower carrier density  $\sim$5.32$\times$10$^{11}$\,cm$^{-2}$~ at 2\,K, which provides direct evidence of protected topological surface states in the \emph{R3m} phase. The Hall conductivity shows a transformation from one band to two-band behavior across the structural transition, thus providing experimental evidence for the degeneracy lifting of bulk valence bands below the cubic symmetry breaking point which is consistent with recent band structure calculations. The overall results indicate that magneto-transport studies can distinctly probe the surface and bulk sensitive properties of SnTe, and can also track the band-splitting of degenerate bands at the Fermi level across the cubic-to-rhombohedral (\emph{R3m}) transition.
\end{abstract}

\maketitle

\section{introduction}
Topological crystalline insulators (TCI) with topological protection due to crystalline symmetries instead of time-reversal symmetry have attracted significant interest in recent times \cite{Hsieh2012,Takahashi2012,Tanaka2013,Assaf2014,Dybko2017,Okazaki2018,Wei2019,Kaloni2019,Chang2019}. SnTe is the first TCI which was predicted to exhibit nontrivial topological surface states (TSS) on (100), (110), (111) planes, with an even number of Dirac cones arising from mirror symmetries with respect to (110) mirror planes \cite{Hsieh2012}. Soon after the original theoretical prediction, the Dirac cone surface states in the electronic structure of SnTe were experimentally confirmed by angle-resolved photoemission spectroscopy (ARPES) experiments \cite{Takahashi2012,Tanaka2013}. 

Extensive studies on SnTe have been carried out much before its topological character was recognized and it is known to exhibit several interesting properties \cite{Brebrick1963prb,Brebrick1963,Brebrick1971,Allgaier1972,Iizumi1975,Kobayashi1976}. It undergoes a displacive/ferroelectric transition coupled to a cubic to rhombohedral structural distortion as a function of temperature \cite{Kobayashi1976,ONeill2020}. This transition depends on the hole carrier density (p) and it was shown that the transition temperature T$_s$ can vary from T$_s$ $\sim$ 100\,K for p $\sim$1$\times$10$^{20}$\cc, which reduces to T$_s$ $\sim$ 30\,K for  p $\sim$8$\times$10$^{20}$\cc and vanishes for p $>$ 1$\times$10$^{21}$\cc \cite{Kobayashi1976}. Early studies showed that the composition stability limits for Sn$_{1-x}$Te$_{x}$ correspond to a very small x-range with x = 0.50015 to 0.5004 for Sn-saturation grown at 400\textcelsius~ and 500\textcelsius, respectively. On the other hand, for Te-saturation, the values of x vary from x = 0.5070 to nearly 0.5074 at 356\textcelsius~ and 400\textcelsius, respectively \cite{Brebrick1963,Brebrick1971}. While this indicates that SnTe is always Te-rich; the hole conductivity can arise from either Sn$^{2+}$ vacancies or Te$^{2-}$ interstitials. An early analysis of composition with density concluded that SnTe showed 85\,\% Sn vacancies with 15\,\% Te interstitials \cite{Brebrick1963}, but a conclusive structural analysis based on line shapes due to diffuse Huang scattering \cite{Dederichs1973,Bhagavannarayana2008,Huang1947} from vacancies or interstitials has not been reported to date. Given this uncertainty of the role of defects and its relation with the hole conductivity, several groups have also investigated thin films of SnTe \cite{Assaf2014,Okazaki2018,Wei2019}. These studies could address interesting aspects of the topological nature of the transport properties such as : (i) weak antilocalization \cite{Assaf2014} for films with low (p* $\sim$8$\times$10$^{18}$\cc) and high (p* $\sim$1$\times$10$^{21}$\cc) hole carrier density, (ii) observation of SdH oscillations and its relation to Rashba splitting of a specific Fermi surface in the electronic structure \cite{Okazaki2018}, and (iii) linear magnetoresistance  behavior \cite{Wei2019}. However, these studies did not distinguish the role of surface and bulk sensitive transport properties in relation to the hole carrier density and defects, or the role of the structural transition in the magneto-transport properties of SnTe films.

In general, for any material, the surface properties in transport measurements are expected to get masked by the bulk carriers. Since SnTe is a heavily self-doped p-type material, it is not easy to separate out the role of surface and bulk sensitive transport properties in SnTe. To our knowledge, only one study \cite{Dybko2017} has reported on the surface and bulk sensitive aspects of the transport properties of as grown SnTe bulk crystals. From a well-designed set of experiments, Dybko et al. have investigated the TSSs and Berry phase of SnTe using SdH oscillations. Furthermore, they also showed that the dHvA oscillations in magnetization matches well with calculations assuming the bulk p-type hole carrier density playing the role of pinning the Fermi energy of the system. 

According to Boltzmann transport theory for conducting materials, the magnetoresistance increases quadratically at low fields and saturates at higher fields. In recent reports on topological insulators \cite{Kumar2018}, Dirac semimetals \cite{Feng2015,Dasoundhi2021}, Ag$_{2\pm\delta}$Se \cite{Hu2007} etc., linear magnetoresistance (LMR) has been observed at high fields. The origin of LMR can be explained by classical or quantum theory. For example, the large unsaturated linear MR in Ag$_{2\pm\delta}$Se \cite{Hu2007}, Bi$_2$Se$_3$ \cite{Kumar2018}, semimetal Sb \cite{Dasoundhi2021} and Cu$_{2-x}$Te \cite{Sirusi2018} etc. has been explained by the classical model by taking into account mobility fluctuations associated with scattering of charge carriers from spatial inhomogeneities.  On the other hand, the LMR observed in topological insulator nano-structures \cite{Wang2012} and graphene \cite{Friedman2010} could only be explained by the quantum model. A giant LMR has also been reported in SnTe hetero-structures \cite{Wei2019}. However, the origin of large LMR in SnTe single crystals and its relation with the classical or quantum model still remains an open question.

On the theoretical front, density functional theory (DFT) based band structure calculations for the \{111\} surface of SnTe have suggested that the characteristic TCI properties are retained in the \emph{R3m} phase when lattice distortion is small \cite{Plekhanov2014}. For the \{100\} surface too, while the rhombohedral distortion breaks the mirror symmetry (100) plane, resulting into two Dirac cones, it is still expected to retain the TCI phase \cite{Hsieh2012,Takahashi2012}. In SnTe, the fundamental bandgap forms at the L points of the cubic Fermi surface (FS). The band structure calculations and ARPES studies in SnTe have predicted the presence of hole pockets at L points of FS for low carrier density, and on increasing carrier density above $\sim$1$\times$10$^{20}$\cc, these pockets get joined by tubes \cite{Littlewood2010}. A recent DFT calculation for the \emph{R3m} structure has revealed the formation of two nondegenerate bands due to cubic symmetry breaking and spin-orbit splitting, but this band splitting could not be identified through ARPES \cite{ONeill2020} measurements.  The Rashba splitting of bulk valence bands due to rhombohedral distortion has also been suggested for the SnTe (111) epitaxial films \cite{Plekhanov2014}. But such a band splitting has not been confirmed for the SnTe (100) plane. More importantly, the evolution of the magneto-transport properties as a function of temperature across the cubic (\emph{Fm-3m}) to rhombohedral (\emph{R3m}) transition has also not been explored experimentally. 

In this work, we have grown SnTe (100) single crystals, characterized the structure and performed detailed magneto transport measurements. We quantify the composition of as grown crystals to be SnTe$_{1.02}$ and identify the presence of Te interstitials from a rocking curve analysis. The electrical resistivity of SnTe$_{1.02}$ single crystal shows evidence for the cubic-to-rhombohedral transition at T$_{s}$ $\sim$ 64\,K and a carrier density of $\sim$1.8$\times$10$^{20}$\cc at 77\,K. We observe a non-saturating positive LMR $\sim$42\,\% at 5\,K and 8\,T field originating from conductivity fluctuations and presence of TSS in the \emph{R3m} phase from SdH oscillations. The LMR is shown to follow the classical Parish-Littlewood model of conductivity fluctuations arising from macroscopic inhomogeneity of tellurium interstitial atoms. From detailed Hall conductivity analysis, we track the evolution of valence band splitting across the cubic to rhombohedral transition and confirm the presence of two non-degenerate bands in the \emph{R3m} phase. 

\section{experimental methods}
SnTe single crystal was grown by modified Bridgman technique. Stoichiometric ratio of Tin and Tellurium were sealed in a quartz tube in the vacuum of $\sim$10$^{-6}$ Torr. The mixture was heated above 800\textcelsius~ followed by natural cooling to room temperature. The detailed characterization confirms the fcc structure having space group \emph{Fm-3m}. A flake from the crystal was mechanically cleaved and all the structural and transport characterizations were done on the same flake. X-ray diffraction (XRD) was performed on a Bruker D8 Advance X-Ray Diffractometer. The $\omega$-scan was done on a Bruker D8 high-resolution X-ray Diffractometer with a Cu-K$_\alpha$ source. Energy Dispersive X-ray (EDX) spectroscopy was carried out on FEI made Nova Nano field emission scanning electron microscopy (FESEM). Resistivity and Hall measurements were done using the AC transport option of the 9\,T PPMS down to liquid Helium temperatures by four probe and five probe method, respectively. Electric contacts were made using indium on the structurally characterized flake of dimensions 2.7$\times$2$\times$0.5 mm$^3$. Magnetic field is applied perpendicular to the sample surface in $<$100$>$ direction for magnetoresistance and Hall measurements.

\section{Results and discussion}
\subsection{Structural characterization}
\begin{figure*}
	\centering
	\includegraphics[width=0.9\linewidth]{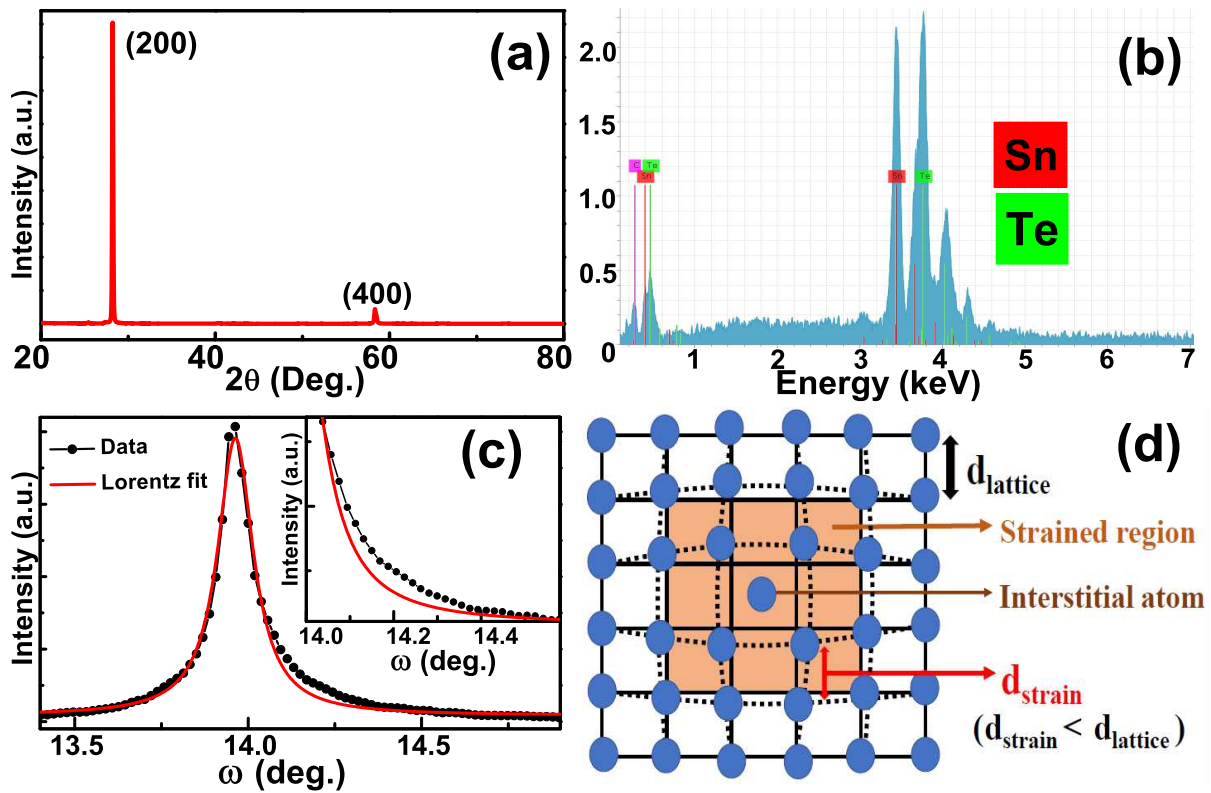}
	\caption{(a) X-ray diffraction of cleaved crystal showing peaks along (200) plane.  (b) EDX spectrum of SnTe shows a small deviation from stoichiometry. (c) Rocking curve of (200) plane (black) along with Lorentzian curve fitting (red line). Inset: Zoomed view showing asymmetry on the right side of rocking curve. (d) Schematic diagram of strained lattice having interstitial defects.}
	\label{fig-1}
\end{figure*}

The out of plane X-ray diffraction (XRD) profile is shown in Fig.\,1(a). The XRD pattern shows the \{200\} family of planes in SnTe single crystal. The crystal showed a lattice parameter of 6.3197(3) $\AA$ consistent with a Te-rich cubic phase \cite{Brebrick1971}. The EDX spectrum of the same flake is shown in Fig.\,1(b) and the stoichiometry of grown crystal is found to be Sn:Te :: 1 :1.02. Fig.\,1(c) shows the $\omega$-scan (rocking curve) of (200) plane. The pattern shows a single peak signifying the high quality of the crystal without structural grain boundaries. However, the full width at half maximum (FWHM) is 0.091\textdegree, which is larger than the standard Silicon sample measured on the same instrument \cite{Kumar2017} and indicates the presence of point defects in the grown crystal. It is known that two kinds of point defects are formed during the growth of SnTe crystals viz. vacancy and interstitial defects \cite{Brebrick1963,Brebrick1971}.  In interstitial defects, the lattice around the interstitial atoms gets compressed and the interplanar spacing (d) decreases (d$_{strain}$ $<$ d$_{lattice}$). This leads to an asymmetric diffuse Huang scattering \cite{Dederichs1973,Bhagavannarayana2008,Huang1947} which causes a broadening on the higher angle side of Bragg peaks. A pictorial representation of an interstitial defect in the lattice is shown in Fig.\,1(d). In contrast, for vacancy type defects, tensile stress occurs around the defects and asymmetric diffuse Huang scattering is expected on the lower angle side of Bragg peaks \cite{Dederichs1973,Bhagavannarayana2008,Huang1947}. As seen from Fig.\,1(c), the right-hand side of the rocking curve of our crystal is slightly broadened. We have fitted the rocking curve to the Lorentzian function (red curve), which clearly shows a small asymmetric scattering intensity on the higher angle side of the Bragg peak indicating the presence of interstitial type of defects in our crystal. Zoomed view of asymmetry on the right side is shown in the inset of Fig.\,1(c).  EDX and Rietveld refinement results show the presence of excess tellurium atoms and also support the existence of interstitial defects. The observation of asymmetric diffuse Huang scattering on the higher angle side of the Bragg peak and presence of excess tellurium atoms observed in EDX and Rietveld refinement indicate that the p-type charge carriers ($\sim$10$^{20}$\cc obtained from Hall measurements, discussed below) can be attributed to the excess interstitial tellurium atoms.
 
\subsection{Transport results and discussions}
The temperature dependent longitudinal resistivity (\rxx) measurements down to 5\,K is shown in Fig.\,2, exhibiting a metallic behavior throughout the measured temperature range. The residual resistivity ratio (RRR= $\rho$$_{250\, K}$⁄$\rho$$_{5\,K}$ is found to be 6. A hump like behavior at 64\,K is observed in the derivative of resistivity vs. temperature (T) curve, as shown in Fig.\,2. This hump like anomaly is a signature of a structural transition, as reported earlier \cite{Kobayashi1976,ONeill2020}. In SnTe, this structural phase transition from cubic (\emph{Fm-3m})-to-rhombohedral (\emph{R3m}) phase is observed below 100\,K and is reported to vanish for carrier density (p$_h$) above $\sim$1.3$\times$10$^{21}$\cc \cite{Kobayashi1976}. In our case p$_h$ is $\sim$1.8$\times$10$^{20}$\cc at 77\,K and the structural deformation appears to be weak, as the signature of transition is pronounced only in the derivative curve. 
 
\begin{figure}
	\centering
	\includegraphics[width=0.7\linewidth]{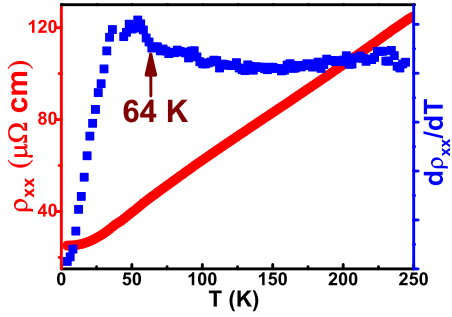}
	\caption{Temperature dependent resistivity (\rxx) curve in the absence of magnetic field (red) and its derivative (d\rxx)⁄dT   curve (blue) with temperature showing a kink at 6\,K.}
	\label{fig:2}
\end{figure}

Fig.\,3(a) shows isothermal magnetic field dependent magnetoresistance at different temperatures where magnetic field is applied perpendicular to the current direction (schematic diagram is shown in the inset of Fig.\,3(a)). A non-saturating positive magnetoresistance ($MR~= \dfrac{\bigtriangleup\rho(B)}{\rho(0)} = \dfrac{\rho(B)-\rho(0)}{\rho(0)}$) is observed where MR increases to 42\,\% with decreasing temperature down to 5\,K and 8\,T field. Isothermal MR shows a crossover from quadratic to linear dependence with field. One such MR curve at 5\,K is shown in Fig.\,3(b) with respective quadratic and linear fittings. According to the Boltzmann electronic transport theory, MR is proportional to ($\mu$B)$^2$ for low fields ($\mu$B $<$ 1) and for closed Fermi surface it saturates at high fields ($\mu$B $>>$ 1) \cite{Singleton2001}. In our case MR varies quadratically at low fields while it becomes linear and non-saturating at higher fields, which cannot be understood by Boltzmann theory.

\begin{figure}
	\centering
	\includegraphics[width=0.7\linewidth]{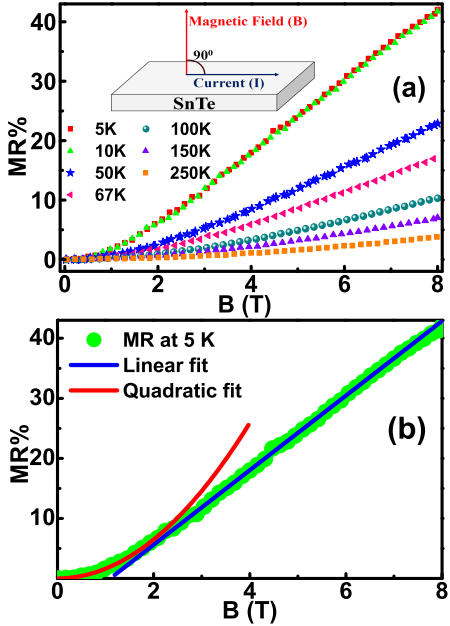}
	\caption{(a) Field dependent isothermal magnetoresistance data at different temperatures. Inset: Schematic diagram of applied magnetic field and current direction on SnTe single crystal. (b) MR\% at 5\,K graph showing quadratic and linear dependence of magnetic field.}
	\label{fig:3}
\end{figure}

 The origin of linear MR is based on two theories viz. classical and quantum. As proposed by Abrikosov \cite{Abrikosov2003}, LMR due to a quantum origin occurs in materials having a linear dispersion relation, only when a single Landau level (LL) is filled; whereas, in the classical case, according to the Parrish and Littlewood (PL) criterion, LMR relates with the mobility fluctuations, arising due to scattering of charge carriers from spatial inhomogeneities \cite{Hu2007,Parish2003}. The PL criterion implies the LMR has a direct dependence on mobility i.e., MR $\propto <\mu>$ or $<\Delta\mu>$  and crossover field B$_c$ $\propto <\mu>^{-1}$ or $<\Delta\mu>^{-1}$ for weak or strong mobility disorder regimes respectively, where  $<\mu>$ is the average mobility and $\Delta\mu$ is the mobility disorder. The crossover field B$_c$ is defined as a field where MR changes from quadratic field dependence to a linear behavior. 

\begin{figure}
	\centering
	\includegraphics[width=.7\linewidth]{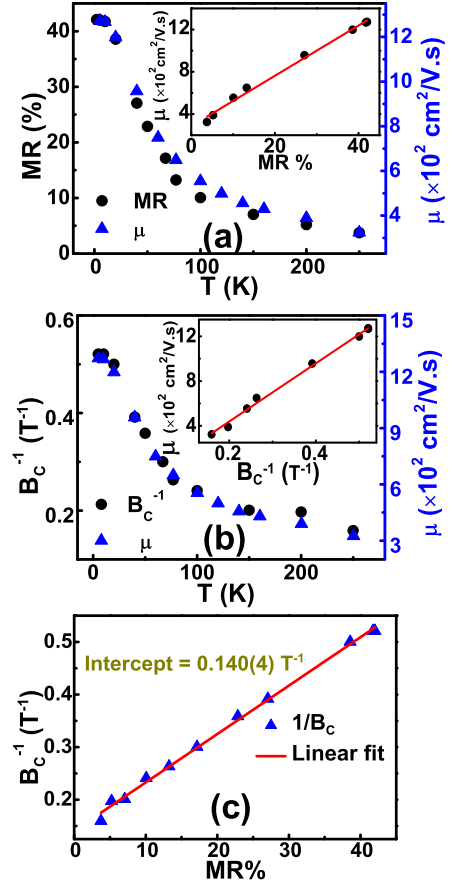}
	\caption{(a) Temperature dependence of MR\% and mobility scaling. Inset: MR\% vs. mobility showing a direct dependence on each other with a linear fit to the data (red color).  (b) Dependence of mobility and B$_c^{-1}$ with temperature, Inset: Direct dependence of mobility with B$_c^{-1}$. Straight line (red color) shows the linear fit to the data. (c) MR\% vs B$_c^{-1}$ plot showing a linear fit (red color) with an intercept 0.14 T$^{-1}$.}
	\label{fig:4}
\end{figure}

In inhomogeneous systems like silver chalcogenides Ag$_{2\pm\delta}$Se and Ag$_{2\pm\delta}$Te \cite{Hu2007}, the mobility fluctuations are explained as a consequence of distortion in current paths due to microscopic defects caused by inhomogeneous distribution of silver atoms. In our SnTe crystal, inhomogeneity is present in the form of tellurium interstitial defects and thus, it is appropriate to investigate the PL conditions for our system. Fig.\,4(a) shows the scaling of MR and mobility with temperature and the inset shows the linear dependence of MR\% with mobility. Similar behavior of B$_c^{-1}$ and mobility with temperature is also observed as shown in Fig.\,4(b), and B$_c^{-1}$ is found to be linearly dependent on mobility as shown in the inset of Fig.\,4(b). The direct proportionality of MR and B$_c^{-1}$ with mobility satisfy the conditions of the weak mobility fluctuation limit of PL model. Hence the observed LMR in SnTe crystal is found to have a classical origin. This can be further analyzed by the intercept of MR Vs B$_c^{-1}$ plot \cite{Sirusi2018} shown in Fig.\,4(c). The observed intercept is almost zero (=0.14\,T$^{-1}$) in our case, which further reconfirms the classical origin of linear MR in single crystal SnTe. The distance between two interstitial tellurium atoms ($\sim$0.102\,$\mu$m) and the carrier mean free path ($\sim$0.14\,$\mu$m) are comparable, suggesting the scattering of charge carriers due to interstitial Tellurium defects are the source of mobility fluctuations leading to LMR in our system.

\begin{figure}
	\centering
	\includegraphics[width=.7\linewidth]{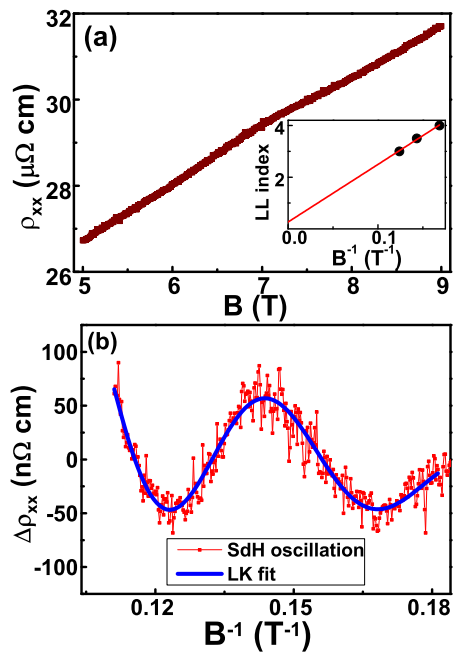}
	\caption{(a) Field dependence of longitudinal resistivity (\rxx) at 2\,K showing SdH oscillations above 5\,T. Inset: LL fan diagram. (b) SdH oscillations along with the LK fit (blue colour solid line).}
	\label{fig:5}
\end{figure}

Next, as shown in Fig.\,5(a), the longitudinal resistivity (\rxx) as a function of applied field reveals SdH oscillations at 2\,K, for applied fields above 5\,T. The oscillatory behavior is clearly observed in $\Delta$\rxx~ as a function of B$^{-1}$, as shown in Fig.\,5(b). The Landau level (LL) fan diagram is plotted in the inset of Fig.\,5(a). The frequency and intercept values obtained from the LL fan diagram are 22±1\,T and 0.2(2) respectively. Oscillations are also analyzed by the Lifshitz-Kosevich (LK) equation. The fitted curve is shown in fig.\,5(b) providing frequency (f)= 22.3(1)\,T, $\gamma$= 0.24(2) and Berry phase ($\beta$)= 0.85$\pi$ \cite{Kumar2015} which is close to $\pi$, confirming the existence of a Berry phase of $\pi$ in our crystal.

In earlier reports, the oscillation frequencies in the bulk SnTe crystals are found to be in the range of $\sim$137-320\,T \cite{Dybko2017,Okazaki2018,ONeill2020,Burke1965,Savage1972}, while the values of oscillation frequencies for topological surface states of SnTe (100) plane is $\sim$26\,T \cite{Dybko2017} and for (111) plane $\sim$10$^{-14}$\,T \cite{Taskin2014}.  In our case for SnTe (100) plane, the oscillation frequency is 22\,T as obtained from the LL fan diagram and the LK fits, which lies close to the values reported for the surface states of the SnTe (100) case. The carrier density of the surface states estimated from the frequency of SdH oscillations is n$_s^{2D}$ $\simeq$~5.32$\times$10$^{11}$\,cm$^{-2}$. The obtained Fermi wave vector (k$_F$) 0.0258 $\AA$$^{-1}$ is also in agreement with ARPES results \cite{Takahashi2012}. The presence of $\pi$ Berry phase and significantly low values of carrier density obtained from the SdH oscillations in our crystal confirms that these low frequency oscillations are emerging from the topological surface states of SnTe (100). 

From theoretical calculations, it has been predicted for the SnTe (100) plane that the rhombohedral distortion breaks one mirror symmetry out of two mirror symmetries with respect to the (110) plane. Consequently, two Dirac cones will be formed but the topological nature due to crystalline symmetry will remain intact \cite{Hsieh2012,Takahashi2012}.  The TCI state in \emph{R3m} phase is also predicted from DFT calculations for the SnTe (111) plane \cite{Plekhanov2014}. As seen from resistivity data, our crystal undergoes a phase transition from cubic to rhombohedral phase at 64\,K. Hence, the existence of TSS at 2\,K from the above analysis on quantum oscillations provides the first experimental evidence of TSS in the rhombohedral phase.

Fig.\,6(a) shows the field dependence of Hall resistivity at different temperatures above the structural transition temperature T$_s$. Hall resistivity is found to be linear above T$_s$ (64\,K) and becomes nonlinear below this temperature as shown in Fig.\,6(b). The linear behavior of Hall resistivity signifies the single band conduction while the nonlinear behavior indicates the presence of more than one band conduction. The two band conduction in SnTe could be due to three possibilities: (i) from surface and bulk band contributions, (ii) from two bulk valence bands forming at L and $\varSigma$ points of the Fermi surface, and (iii) due to the lifting of degeneracy of bulk valence band in \emph{R3m} phase as theoretically predicted by O’Neil et al. \cite{ONeill2020} The first two possibilities are independent of the structural phase transition, while the degeneracy lifting occurs at the structural phase transition.  

\begin{figure*}
	\centering
	\includegraphics[width=0.7\linewidth]{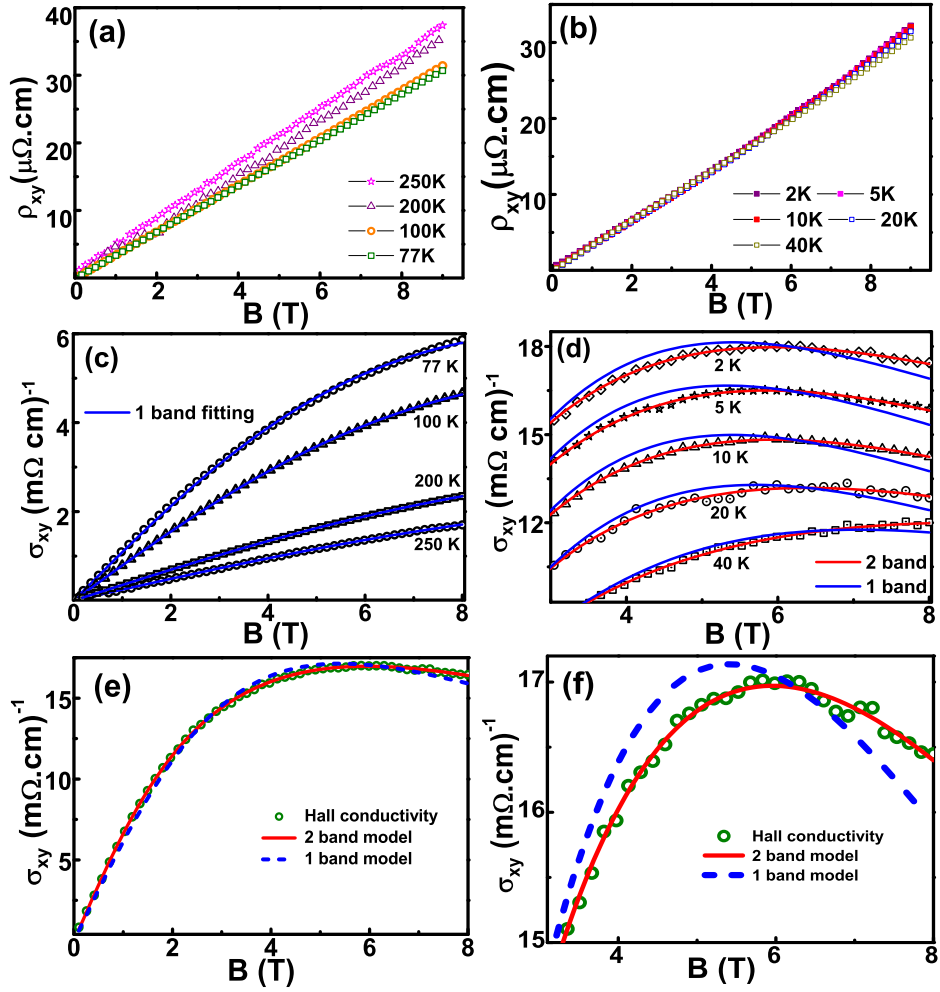}
	\caption{(a) Linear Hall resistivity (\rxy) at temperatures above T$_s$ (b) Nonlinear \rxy at temperatures below T$_s$ (c) Field dependent Hall conductivity (\cxy) curves (symbols) with respective single band fitting (blue) for temperatures above T$_s$. (d) \cxy~ for temperatures below T$_s$ with respective single band (blue) and two-band (red) fitting. (e) Field dependent \cxy~ at 2\,K. Blue dotted line shows single band fitting and red solid line shows two band fitting to the data. (f) \cxy~ at higher fields showing divergence from single band fit at 2\,K.}
	\label{fig:6}
\end{figure*}

We have carried out a careful analysis of the Hall conductivity $( \sigma_{xy} = \dfrac{\rho_{xy}}{\rho_{xy}^2+\rho_{xx}^2} )$ in order to confirm the two band conduction mechanism. The Hall conductivity data was fitted using a least-squares method to the two band model using the equation: 
\[
\sigma_{xy} = \dfrac{p_1e\mu_1^2B}{1+\mu_1^2B^2} + \dfrac{p_2e\mu_2^2B}{1+\mu_2^2B^2}
\]
where p$_{1,2}$  and $\mu_{1,2}$  are carrier densities and mobilities of band-1 and 2, respectively.  Figures 6\,(c) and (d) show the Hall conductivity (\cxy) curves overlaid with one band fitting curves for several temperatures above T$_s$ and two band fitting curves for temperatures below T$_s$, respectively. At low temperatures below the structural transition, deviation from the single band model fit is clearly visible at higher fields whereas the Hall conductivity fits best with the two band model, as shown by red lines in Fig.\,6(d). For clarity, the Hall conductivity data measured at T = 2\,K is shown in Fig.\,6(e) for both, the one band and two-band fits. The zoomed view at higher fields is shown in the Fig.\,6(f) to emphasize the deviation from the one band fit and a reasonably good match to the two band model. It is noted that in a recent study reporting the Hall conductivity under applied pressures of 6 kbar and 10\,kbar, it was shown that the data could be fitted with a one band model and a two band model did not improve the fit. However, the authors also showed that the structural distortion defining the rhombohedral distortion, namely, the polar displacement was strongly suppressed and the system was very close to the cubic phase for an applied pressure of 6\,kbar, and for 10\,kbars, SnTe was well into the cubic phase \cite{ONeill2020}. 

Temperature dependence of carrier concentration and mobility obtained from the two band and single band fittings are plotted in Fig.\,7(a) and 7(b) respectively. Both the bands have holes as charge carriers in this system. The values of carrier densities and mobilities for band-1 are $\sim$1.2$\times$10$^{20}$\cc, $\sim$1400\,cm$^2$/V.s whereas for band-2 are $\sim$2$\times$10$^{19}$\cc and $\sim$3000\,cm$^2$/V.s respectively. In order to confirm the reliability of fitting parameters the zero field conductivity is calculated from the obtained parameters and compared with the observed value of conductivity. The calculated and measured conductivity in zero field are consistent with each other as shown in Fig.\,7(c). The calculated zero field resistivity $\sim$2.78$\times$10$^{-5}$  $\Omega$.cm also matches with the measured residual resistivity ($\sim$2.5$\times$10$^{-5}$ $\Omega$.cm) of the sample, further validating the two band model in our system.

The large mobility of the second band from the Hall conductivity fitting may be presumed to arise from surface states. However, the carrier density of this band ($\sim$2$\times$10$^{19}$\cc) is much larger than the carrier density of the surface states estimated from the frequency of the SdH oscillations (n$_s^{2D}$ $\simeq$ 5.32$\times$10$^{11}$\,cm$^{-2}$~), ruling out the possibility of the second band representing the surface states.  Further if the surface bands are significantly contributing to the conductivity, the resistivity of the sample should be $\sim$67.11 $\Omega$.cm as calculated from carrier concentration obtained from SdH oscillations, while the actual resistivity obtained is much smaller ($\sim$2.5$\times$10$^{-5}$ $\Omega$.cm). This observation signifies that the surface transport is not the determining factor for the magneto transport behaviour of our system. This also confirms that the classical LMR in our system is a direct consequence of mobility fluctuations of bulk carriers rather than surface carriers.

\begin{figure}
	\centering
	\includegraphics[width=0.7\linewidth]{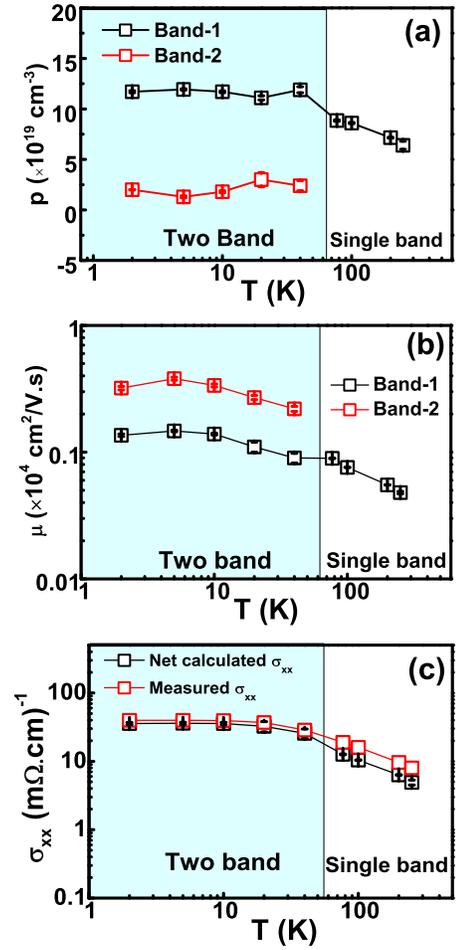}
	\caption{(a) Carrier density and (b) mobility of two bands, derived from Hall conductivity. (c) The estimated conductivity \cxx (black) plotted along with the measured \cxx (red) for 0\,T magnetic field.}
	\label{fig:7}
\end{figure}

According to earlier reports of MR \cite{Allgaier1972} and thermopower measurements \cite{Brebrick1963prb}, SnTe can have contributions from two bulk valence bands (VB) in the vicinity of the L and $\varSigma$ points in the Brillouin zone, but the major contribution is from the valence bands near the L point. However, the ARPES results on SnTe crystal having carrier density $\sim$3$\times$10$^{20}$\cc have revealed that the Fermi level barely cuts the top of the valence band at $\varSigma$ point, indicating an insignificant number of carriers participating from the $\varSigma$ valence bands. The carrier concentration estimated from the k$_F$ $\simeq$ 0.042 $\AA$$^{-1}$ value obtained from ARPES results \cite{ONeill2020} at $\varSigma$ point is found to be $\sim$10$^{18}$\cc, a value which is much smaller than the values obtained from our Hall results. Our carrier density from Hall measurements for the second band is an order of magnitude larger. Hence, we discard the possibility of the valence band at $\varSigma$ point being the second band in our case. Consequently, we rule out the possibility of having distinct VB$_L$ and VB$_\varSigma$ bands in our transport mechanism and hence the second band neither corresponds to surface carriers nor to VB$_\varSigma$ carriers.

Based on a beating pattern in the SdH oscillations, Okazaki et al. \cite{Okazaki2018} have suggested that the two components in SdH oscillations observed in \emph{R3m} phase of SnTe (111) films most likely arise from a Rashba splitting of the bulk valence bands. Similar predictions of band splitting were also made by band structure calculations and is attributed to lifting of degeneracy in \emph{R3m} phase \cite{ONeill2020}. In our case, the nonlinearity in Hall resistivity appears below the structural phase transition at 64\,K, implying that the non-linearity arises from a lifting of degeneracy owing to the splitting of bulk valence bands. In our crystal, the mobility of charge carriers in the second band is comparable with the reported mobility ($\sim$2570$\pm$50 cm$^2$/V.s) in \emph{R3m} phase \cite{ONeill2020}. Therefore, our experimental observations of the presence of two bands in the rhombohedral phase and comparable mobility of charge carriers in the two bands confirm that they originate from the degeneracy lifting of the bulk valence bands due to cubic symmetry breaking. 

\begin{figure}
	\centering
	\includegraphics[width=0.7\linewidth]{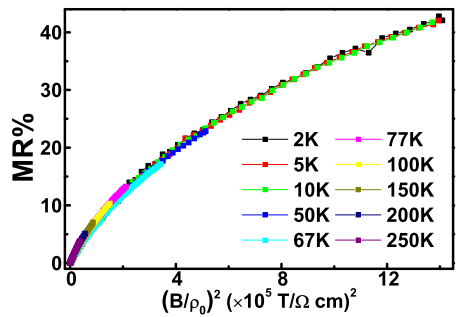}
	\caption{Kohler’s plot showing the magnetoresistance as a function field for all measured temperatures get merged into a single curve.}
	\label{fig:8}
\end{figure}

The temperature dependence of mobility of both the bands are shown in Fig.\,7(b) and they follow a remarkably similar trend indicating the similar temperature dependence of carrier scattering time. This suggests a similar scattering mechanism is operative in both the bands. We investigated this point further by attempting a Kohler scaling analysis of the data. According to Kohler’s rule, plots of MR versus (B/$\rho_0$)$^2$ merge into a single curve for all temperatures if the charge carriers are from a single band, or if the carriers from two bands have the same temperature dependence of scattering rate. As seen from Fig.\,8, this scaling is realized in our case as the Kohler plots at all temperatures collapse into a single curve. This confirms a single scattering mechanism of all the carriers participating in transport properties. Interestingly, the merger of all the curves below and above the structural transition temperature signify that the scattering mechanism is not affected even across the structural phase transition. This indicates that the two bands in the low temperature phase originate from a single band in the high temperature phase. Therefore, the presence of two bands in our system having same scattering mechanism of charge carriers in the rhombohedral phase, and also for the single band in the cubic phase, provides a strong experimental evidence of degeneracy lifting of the bulk L-pocket bands in the \emph{R3m} phase. 

\begin{figure}
	\centering
	\includegraphics[width=0.7\linewidth]{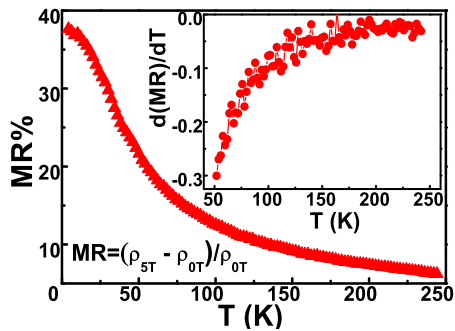}
	\caption{Variation of MR\% (obtained using 5\,T and 0\,T resistivity data) as a function of temperature. Inset: Derivative of MR\% as a function of temperature.}
	\label{fig:9}
\end{figure}

The effect of structural transition on MR can also be explored by temperature dependence of MR as shown in Fig.\,9, and its derivative is shown in the inset to Fig.\,9. Both the curves show a smooth change with temperature, unlike the hump observed in \rxx(T) curve, which indicates that the structural transition does not affect the magnetoresistance, which is in accordance with our Kohler analysis.

\section{conclusion}
In summary, we have performed a detailed magneto transport study of single crystal SnTe (100) which shows a structural phase transition at 64\,K. A non-saturating linear magnetoresistance of 42\% at 5\,K and 8\,T is observed and it exhibits a crossover from a quadratic to linear dependence as a function of magnetic field. The LMR matches well with the classical PL model governed by scattering of charge carriers from the inhomogeneities due to Te interstitial defects. The SdH quantum oscillation analysis asserts the presence of nontrivial TSS in \emph{R3m} phase even after the cubic symmetry breaking. Nonlinearity in Hall resistivity at low temperature confirms the presence of two bands in the rhombohedral phase which is well corroborated by the recent theoretical prediction of lifting of band degeneracies due to structural distortion.

\begin{acknowledgments}
We thank Mukul Gupta and Layant Behera for XRD measurements; V. R. Reddy and Anil Gome for HRXRD measurements; R. Venkatesh and S. Poddar for EDX measurements. D. K. acknowledges research grant from SERB India for Early Career Research Award (ECR/2017/003350).
\end{acknowledgments}


%

\end{document}